# AGU Advances

**Correspondence to:**
S. Terpstra,
s.terpstra@uu.nl

**Author Contributions:**
**Conceptualization:** Sjoerd Terpstra, Henk A. Dijkstra, Anna S. von der Heydt
**Data curation:** Sjoerd Terpstra
**Formal analysis:** Sjoerd Terpstra
**Funding acquisition:** Anna S. von der Heydt
**Investigation:** Sjoerd Terpstra
**Supervision:** Robbin Bastiaansen, Henk A. Dijkstra, Anna S. von der Heydt
**Visualization:** Sjoerd Terpstra
**Writing – original draft:** Sjoerd Terpstra, Swinda K. J. Falkena, Anna S. von der Heydt
**Writing – review & editing:**
Sjoerd Terpstra, Swinda K. J. Falkena, Robbin Bastiaansen, Sebastian Bathiany, Henk A. Dijkstra, Anna S. von der Heydt

# Assessment of Abrupt Shifts in CMIP6 Models Using Edge Detection

Sjoerd Terpstra[1,2] , Swinda K. J. Falkena[1] , Robbin Bastiaansen[1,3] , Sebastian Bathiany[4,5] , Henk A. Dijkstra[1,2] , and Anna S. von der Heydt[1,2]

[1]Institute for Marine and Atmospheric Research Utrecht, Utrecht University, Utrecht, the Netherlands, [2]Centre for Complex Systems Studies, Utrecht University, Utrecht, the Netherlands, [3]Mathematical Institute, Utrecht University, Utrecht, the Netherlands, [4]Earth System Modelling, School of Engineering and Design, Technical University of Munich, Munich, Germany, [5]Potsdam Institute for Climate Impact Research, Potsdam, Germany

**Abstract** Past research has shown that multiple climate subsystems might undergo abrupt shifts, such as the Arctic Winter sea ice or the Amazon rainforest, but there are large uncertainties regarding their timing and spatial extent. In this study we investigated when and where abrupt shifts occur in the latest generation of earth system models (CMIP6) under a scenario of 1% annual increase in $CO_2$. We considered 82 ocean, atmosphere, and land variables across 57 models. We used a Canny edge detection method to identify abrupt shifts occurring on yearly to decadal timescales, and performed a connected component analysis to quantify the spatial extent of these shifts. The systems analyzed include the North Atlantic subpolar gyre, Tibetan Plateau, Amazon rainforest, Antarctic sea ice, monsoon systems, Arctic summer sea ice, Arctic winter sea ice, and Barents sea ice. Except for the monsoon systems, we found abrupt shifts in all of these across multiple models. Despite large inter-model variations, higher levels of global warming consistently increase the risk of abrupt shifts in CMIP6 models. At a global warming of 1.5°C, six out of 10 studied climate subsystems already show large-scale abrupt shifts across multiple models.

**Plain Language Summary** This study investigates abrupt shifts in climate subsystems such as sea ice, monsoon systems, and permafrost, which could have severe impacts on the planet. It quantifies where and when these shifts might occur by analyzing the latest climate subsystems models under a simulation of increasing $CO_2$. Using edge detection—a method to detect abrupt shifts—we identified which subsystems are vulnerable to abrupt shifts and under what levels of global warming. This helps evaluating the risks that specific climate subsystems undergo abrupt shifts under the effect of global warming. At a global warming of 1.5°C—which is a target set by the Paris climate agreement—six out of 10 studied climate subsystems showed large-scale abrupt shifts across multiple models.

## 1. Introduction

Due to increasing levels of greenhouse gases, the climate is changing rapidly, bringing about negative consequences such as loss of biodiversity (Bellard et al., 2012; Vasiliev & Greenwood, 2021), rising sea level (Day et al., 2008), and more numerous extreme events (Seneviratne et al., 2021). Additionally, there is concern that this increase in greenhouse gases could lead to large and sudden changes in the climate system known as abrupt shifts. An abrupt shift is defined as "a change that takes place substantially faster than the rate of change in the recent history of the affected component of a system" according to the most recent IPCC report (Intergovernmental Panel On Climate Change, 2023, p. 202). A related concept is that of a tipping point (Armstrong McKay et al., 2022; Lenton et al., 2008) defined as "a critical threshold beyond which a system reorganizes, often abruptly and/or irreversibly" (Intergovernmental Panel On Climate Change, 2023, p. 202). Crossing a tipping point can lead to an abrupt shift when a system reorganizes in a sudden manner. However, tipping dynamics can also occur over much longer timescales compared to the internal dynamics of these systems—and thus does not have to be abrupt. In this paper, we focus on abrupt shifts in which a system changes faster than expected compared to its recent history.

Abrupt shifts have been found in many simulations using climate models of different complexity (Höning et al., 2023; Swingedouw et al., 2021; van Westen & Dijkstra, 2023). The potential for abrupt shifts is further supported by evidence from paleoclimate, such as the recurring abrupt shifts during Dansgaard-Oeschger events (Boers et al., 2022; Dansgaard et al., 1993). During these events, rapid warming occurs on a timescale of decades, likely linked to abrupt changes in sea ice cover and the Atlantic Meridional Overturning Circulation (AMOC) (Henry et al., 2016; Wunderling et al., 2024). Since such sudden changes can have large impacts on ecosystems and society (Brovkin et al., 2021; Dietz et al., 2021; McNeall et al., 2011), predicting them with good accuracy is crucial.

Yet, there are large uncertainties surrounding abrupt shifts. For example, it is unclear under which conditions—such as the level of global warming—climate subsystems undergo abrupt shifts (Armstrong McKay et al., 2022; Lenton, 2023; Lenton et al., 2008; Wang et al., 2023). Efforts to reduce these uncertainties include the work by Drijfhout et al. (2015) on abrupt shifts in a recent generation of earth system models (CMIP5) (Taylor et al., 2012). In Drijfhout et al. (2015), all CMIP5 models were scanned for abrupt shifts in simulations of potential future climates. The authors found evidence of abrupt shifts in a variety of systems and demonstrated that most of them had a limited spatial extent, with the majority involving ocean-sea ice interactions.

While the results of Drijfhout et al. (2015) advanced our understanding of abrupt shifts in the climate, the downside of their method was that it involved manual steps. To improve the detection of abrupt shifts in climate data, Bathiany et al. (2020) adapted the Canny edge detection method (Canny, 1986)—typically used in computer vision—for usage with climate data. This method detects edges, which are points in space and time where abrupt changes occur. Hence, it is possible to automatically detect abrupt shifts in climate data. The authors used this method to scan the same CMIP5 models as Drijfhout et al. (2015). Both studies reached similar results, with the highest fraction of abrupt shifts detected in polar regions, mostly in the Northern Hemisphere. While Bathiany et al. (2020) validated the edge detection method on the CMIP5 models, their study was limited to a general scan of abrupt shifts where they only counted the number of abrupt shifts on the grid cell level per variable. They did not quantify global warming thresholds at which these shifts occur, nor did they analyze the spatial extent of abrupt shifts.

Recognizing the usefulness of edge detection, we apply it to the current state-of-the-art earth system models (CMIP6) which have not yet been subjected to a comprehensive scan across all major systems that can undergo abrupt shifts (however, several studies have examined specific subsystems (e.g., Parry et al., 2022; Swingedouw et al., 2021)). The CMIP6 models are the latest generation of fully coupled earth system models (Eyring et al., 2016) used as a basis for the IPCC reports. For this study, we consider the standard 1pctCO$_2$ simulation of CMIP6. This is a simulation over 150 years in which the CO$_2$ concentration is increased annually by 1% until the CO$_2$ concentration is four times the preindustrial level at 140 years in the simulation. We use 82 two-dimensional land, ocean, and atmosphere variables such as vegetation, sea ice, and temperature fields (see Table S2 in Supporting Information S1 for the full list). The strength of this study lies in determining the global warming thresholds and spatial extent of abrupt shifts using edge detection on the updated CMIP6 models, which offer a better representation of physical processes compared to the CMIP5 models.

Our analysis is limited to detecting abrupt shifts on decadal timescales or shorter, as the 1pctCO$_2$ simulations in CMIP6 models span only 150 years. In addition, CMIP6 models do not include all components of the climate system that can undergo abrupt changes. For example, the Greenland ice sheet, which can rapidly decline after crossing a tipping point (Robinson et al., 2012), is not dynamically represented in these simulations. Our edge detection method identifies statistically significant abrupt shifts but does not assess the underlying system dynamics. As a result, we cannot determine whether these abrupt changes are regime shifts persistent on centennial—or longer—timescales. or represent qualitative changes in the system, such as those caused by crossing a tipping point.

We investigate abrupt shifts occurring in different climate subsystems whose physical dynamics are well-represented within the CMIP6 models. The structure of our analysis is based on the overview provided by the recently published Global Tipping Points Report (GTP-report) (Lenton, 2023). This extensive assessment synthesizes current evidence for tipping dynamics in various earth system components. It ranks climate subsystems as tipping elements, unclear tipping elements, or not tipping elements based on their plausibility of possessing tipping points. For all these systems, it also assesses whether these systems might change abruptly or not—which is the focus of our study. We have used it as a guide for which subsystems to investigate and we discuss our results in light of this GTP-report.

Based on these considerations, we analyze all climate subsystems that are adequately represented in CMIP6 and have the potential for abrupt shifts under the 1pctCO$_2$ scenario (see Text S1.E in Supporting Information S1 for a detailed justification of the included systems). Of the subsystems identified as tipping elements with low to high confidence in the GTP-report (Lenton, 2023), we analyze the North Atlantic subpolar gyre, the Tibetan Plateau, the land permafrost in the Northern Hemisphere, the Amazon rainforest, and the Boreal forests. Of the subsystems identified as unclear tipping elements, we analyze the Indian, West African and South American Monsoons, and the Antarctic sea ice. Finally, for subsystems identified as not tipping elements, we consider the Arctic summer sea ice, the Arctic winter sea ice, and the Barents sea ice. Although these last three subsystems are not classified as tipping elements, they still exhibit abrupt transitions in CMIP5 models (Bathiany et al., 2020; Drijfhout et al., 2015). For each of these subsystems, we analyze a number of key variables which are chosen based on their physical relevance in representing the system and evidence of this literature, as well as their availability for a large number of CMIP6 models.

Since the edge detection method identifies all statistically significant changes in variables regardless of the size of the shift, we need to distinguish minor fluctuations from actual abrupt shifts that occur over a decadal timescale. To this end, we apply an abruptness measure adapted from Bathiany et al. (2020). This measure quantifies the amplitude of the detected shift relative to the typical variance of the variable (measured from a control simulation with $CO_2$ concentration fixed at preindustrial levels). The shift is only classified as abrupt if the amplitude of change is large enough (see Section 2 for more details).

The edge detection identifies abrupt shifts that occur at the grid cell level of the models. To separate between spatially localized and large-scale abrupt shifts within climate subsystems, we apply a connected component algorithm to group neighboring abrupt shifts to identify regions where they occur simultaneously. In the rest of this paper, by an abrupt shift we mean the full connected component experiencing an abrupt shift except when stated otherwise. A large-scale abrupt shift is defined as covering an area of more than a given threshold, unique for each system. We link the large-scale abrupt shifts to the level of global warming at which they occur in the models.

## 2. Materials and Methods

We searched for abrupt shifts in 57 CMIP6 models run under the 1pctCO2 scenario for a total of 82 different two-dimensional variables with monthly frequency (see Table S2 in Supporting Information S1 for the full list of variables and Table S3 in Supporting Information S1 for the full list of models). Simulations were selected based on their availability at the CMIP6 archive of the Earth System Grid Federation (ESGF) (Cinquini et al., 2014).

### 2.1. Key Variables

Apart from the global scan on 82 different variables (of which a large part has been selected from Bathiany et al. (2020)), we focus the analysis of the specific subsystems on a set of key variables that best represent these systems. These variables were chosen based on three criteria: (a) variables that are physically relevant to the system, (b) variables commonly investigated in the literature assessing abrupt shifts, (c) variables that are widely available within the CMIP6 ensemble. The key variables for each subsystem are shown in Table 1.

For the North Atlantic subpolar gyre, mixed-layer depth is ideally used, as it represents convection. However, its large variability makes it challenging to analyze with our methodology. Therefore, other proxies were also included in the analysis. Sea surface temperature was used to assess the state of the subpolar gyre since a collapse of the subpolar gyre can lead to abrupt cooling at the surface (Swingedouw et al., 2021). The collapse of the subpolar gyre can be associated with freshening of the surface layer, which is why sea surface salinity was also included (Swingedouw et al., 2021). Additionally, due to its broader coverage in CMIP6 models, surface air temperature was also included (Swingedouw et al., 2021).

To assess abrupt shifts in the Tibetan plateau, we analyzed variables related to snow cover. The variables snow area percentage, surface snow melt, and snow depth directly represent this. Furthermore, during abrupt changes in snow cover, large changes in temperature are also expected; thus near-surface air temperature was included. Finally, surface upwelling shortwave radiation was used as an indicator of albedo.

For land permafrost, we analyzed soil frozen water content and soil moisture content. The first was included as it serves as a proxy for permafrost (Steinert et al., 2023). The second was previously analyzed by Drijfhout et al. (2015) for CMIP5.

**Table 1**
*Overview of the Analyzed Climate Subsystems*

| | Latitude | Longitude | Key variables | Minimum area (km²) |
|---|---|---|---|---|
| **Subsystems identified as tipping elements** | | | | |
| North Atlantic subpolar gyre | (45, 60) | (−70, −10) | Mixed-layer depth, sea surface salinity, sea surface temperature, surface air temperature | − |
| Tibetan Plateau | (25, 40) | (70, 105) | Snow area percentage, surface snow melt, snow depth, near-surface air temperature, surface upwelling shortwave radiation | $10^5$ |
| Land permafrost | (45, 90) | (−180, 180) | Soil frozen water content, total soil moisture content | $10^6$ |
| Amazon rainforest | (−20, 5) | (−80, −50) | Bare soil percentage area coverage, carbon mass in vegetation, carbon mass in soil pool, carbon mass flux into atmosphere due to fire, total carbon mass flux from vegetation to litter/from vegetation directly to soil, carbon mass flux out of atmosphere due to gross primary production on land, natural grass area percentage, leaf area index, total atmospheric respiration on land as carbon mass flux, tree cover percentage | − |
| Boreal forests | (45, 80) | (−180, 180) | Bare soil percentage area coverage, carbon mass in vegetation, carbon mass in soil pool, carbon mass flux into atmosphere due to fire, total carbon mass flux from vegetation to litter/from vegetation directly to soil, carbon mass flux out of atmosphere due to gross primary production on land, natural grass area percentage, leaf area index, total atmospheric respiration on land as carbon mass flux, tree cover percentage | − |
| **Subsystems identified as unclear tipping elements** | | | | |
| Antarctic sea ice | (−90, −55) | (−180, 180) | Sea ice concentration, sea ice thickness, sea-ice mass change through bottom melting and surface melting, surface upwelling shortwave radiation, near-surface air temperature | $5 \times 10^5$ |
| South American Monsoon | (−75, −35) | (−25, 5) | Precipitation, convective precipitation, latent heat flux | − |
| West African Monsoon | (−20, 30) | (0, 25) | Precipitation, convective precipitation, latent heat flux | − |
| Indian Summer Monsoon | (65, 100) | (5, 35) | Precipitation, convective precipitation, latent heat flux | − |
| **Subsystems identified as not tipping elements** | | | | |
| Arctic summer and winter sea ice | (60, 90) | (−180, 180) | Sea ice concentration, sea ice thickness, sea-ice mass change through bottom melting and surface melting, surface upwelling shortwave radiation, near-surface air temperature | $10^6$ |
| Barents sea ice | (65, 81) | (15, 60) | Sea ice concentration, sea ice thickness, sea-ice mass change through bottom melting and surface melting, surface upwelling shortwave radiation, near-surface air temperature | $2 \times 10^5$ |

*Note.* The table shows the latitude and longitude bounds, the key variables (which are the focus of the analysis), and the minimum area required for an abrupt shift to be considered large-scale for each subsystem ("−" means there is no minimum area threshold). The Amazon rainforest and Boreal forests share the same key variables, as do the Antarctic sea ice, the Arctic summer and winter sea ice, and the Barents sea ice.

To assess abrupt changes in the vegetation of Boreal forests and Amazon rainforest, we chose to include all CMIP6 variables that measure some part of vegetation. These variables were used in the analysis of abrupt shifts of vegetation by Bathiany et al. (2020), Drijfhout et al. (2015), and Parry et al. (2022).

For the monsoon systems we analyzed (convective) precipitation and latent heat flux. Precipitation is the main characteristic of the monsoon. Latent heat flux is closely linked to precipitation (Zeng & Zhang, 2020) and therefore we included it to assess the state of the monsoon systems. Since we only scan CMIP6 variables in the analysis, we did not include any monsoon indices that require computation beforehand (e.g., Zeng & Zhang, 2020).

To detect abrupt shifts in sea ice systems, sea ice concentration and sea ice thickness were analyzed since they are direct measures of the state of the ice. Additionally, sea-ice mass change through bottom melting and surface melting were used to detect abrupt shifts in melt rates. Surface upwelling shortwave radiation was used as an indicator of abrupt shifts in albedo, signaling sea ice loss. Finally, near-surface air temperature, highly dependent on to the presence of sea ice, was included as key variable.

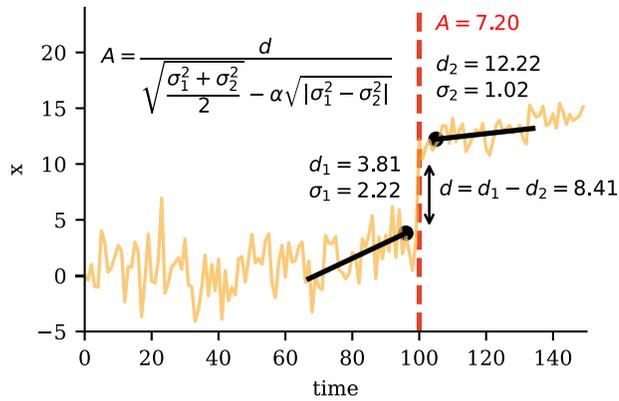

**Figure 1.** Construction of the abruptness measure. The data shown comes from a Gaussian process with a step increase of 10 units at 100 time units where the red dashed line denotes the edge with the corresponding abruptness value $A$. The abruptness is calculated with Equation 1 with $d$ the distance between the two black points $d_1$ and $d_2$ and the standard deviations of the black segments $\sigma_1$ and $\sigma_2$. The asymmetric noise correction term is parameterized by the parameter $\alpha = 0.4$.

### 2.2. Edge Detection

To detect abrupt shifts, we used Canny edge detection (Canny, 1986), as extended by Bathiany et al. (2020) for climate data in many dimensions. Here, we use two spatial dimensions and one temporal dimension. For a full derivation of the underlying method, we refer the reader to Canny (1986). Since we used the edge detection method as implemented by Bathiany et al. (2020), we only provide an overview of the method here. For a complete discussion and implementation details of the edge detector, please refer to Text S1 in Supporting Information S1 and Bathiany et al. (2020). The relevant parameters are given in Table S1 in Supporting Information S1.

The edge detector requires a 1pctCO2 simulation and a pre-industrial control simulation (with pre-industrial CO2 concentration). During preprocessing, the data is regridded to a rectilinear grid, and a Gaussian filter is applied to enhance the signal-to-noise ratio (Bathiany et al., 2020). The edge detector is then applied to each month and to the annual time series of the 1pctCO2 simulation. The edge detector first calculates space-time gradients using the Sobel operator, retaining only locally maximum gradients (non-maximum suppression). Hysteresis thresholding is then applied to identify *strong edges* and discard insignificant events (those below a lower threshold or not connected to a *strong edge*). The pre-industrial control simulation is used as a baseline for internal variance, with hysteresis thresholds set at 95% (upper) and 50% (lower) of the maximum signal.

Abrupt shifts are defined as edges with a high value for the abruptness. To determine abruptness, we used an adjusted Cohen's D as a metric for effect size, replacing the Cohen's D used by Bathiany et al. (2020) and Cohen (1977). Figure 1 illustrates how the abruptness measure $A$ was calculated from the time series. Two segments of the time series, each with a length up to $c_{max}$, were extracted on either side of the edge, with a gap of $c_{trans}$ between them. If either segment was shorter than $c_{min}$, the edge was discarded due to insufficient data for accurate calculation. The abruptness measure is then defined as

$$A = \frac{d}{\sqrt{\frac{\sigma_1^2 + \sigma_2^2}{2} - \alpha\sqrt{|\sigma_1^2 - \sigma_2^2|}}}, \tag{1}$$

with $\sigma_1$ and $\sigma_2$ the standard deviations of the two segments. The term following $\alpha$ is an adjustment to better detect abrupt shifts with large differences in variability before and after the shift. The optimal value of $\alpha$ was determined through various test cases to minimize the number of false positives (see Figure S1 in Supporting Information S1). We set $\alpha = 0.4$ for the analysis.

The abruptness measure quantifies the size of a shift relative to internal variability. We kept only those abrupt shifts for which the abruptness was higher than 4. When the variance before and after the edge is equal, this is equivalent to stating that the abrupt shift is greater than 4 times the standard deviation. This is a relatively strict criterion in order to reduce the likelihood of classifying non-abrupt changes as abrupt shifts. This is the same threshold as used by Drijfhout et al. (2015) and Bathiany et al. (2020).

We conducted a sensitivity analysis on the $\alpha$ parameter of the abruptness measure—in addition to the test cases used to set $\alpha$—by using the detected large-scale abrupt shifts in the CMIP6 models (see Section Connected components for the definition of large-scale abrupt shifts). Specifically, we tested the sensitivity for $\alpha = 0$ which gives the measure used by Bathiany et al. (2020), and $\alpha = 0.4$ which gives the measure used in this study, aimed at improving the robustness of the abruptness measure for variables with highly unequal variance before and after a shift. We selected two variables of two subsystems: the mixed-layer depth in the subpolar gyre (highly unequal variance) and Arctic summer sea ice concentration (relatively low variance due to averaging over large regions). As expected, for the mixed-layer depth, increasing $\alpha$ resulted in more detected abrupt shifts, with the number of models with a significant abrupt shift increasing from 3 to 11 models. In contrast, the number of models with detected large-scale abrupt shifts for Arctic sea ice concentration remained the same at 21 models, with some increase in the total number of large-scale abrupt shifts (meaning more abrupt shifts over the same set of models) from 174 to 195 shifts. This indicates that the influence of $\alpha$ on the detection of abrupt shifts is less strong for variables with more uniform variance than highly unequal variance, as was intended.

### 2.3. Connected Components

To make statements regarding the spatial extent of abrupt shifts and to facilitate comparison with Drijfhout et al. (2015), we applied a connected component algorithm to the edges, allowing us to obtain connected regions that experience an abrupt shift simultaneously. The algorithm consists of two steps. First, small gaps in both the temporal and spatial dimensions are closed, connecting separate events that are relatively close. This step is crucial to avoid identifying numerous disconnected abrupt shifts. Second, the connected components are found by scanning surrounding cells in space and time for adjacent edges, grouping them into cohesive regions. Details of the implementation can be found in Text S1.B1 in Supporting Information S1.

To verify whether the connected components still undergo an abrupt shift, we calculated the spatially weighted mean of the variable in question for these regions. We then performed a one-dimensional edge detection to identify the timing of the abrupt shift (see Text S1.A1 in Supporting Information S1 for the algorithm of the one-dimensional edge detector, which is a simplified version of the three-dimensional edge detector). Using the same abruptness threshold as before, we classified a shift as abrupt if the abruptness exceeds 4. The area of the connected component was then used to determine whether the abrupt shift was large-scale, determined by an area threshold specific to each studied subsystem, as summarized in Table 1.

### 2.4. Onset of Abrupt Shifts Measured by Global Warming Level

To compare the timing of abrupt shifts across models, we used the global warming level at which each shift occurred, rather than the simulation year. This is because models warm at different rates and therefore reach different levels of global warming under the same $CO_2$ concentration. As a result, the simulation year in which the abrupt shift occurs is not a good basis for comparison. Aligning models by global warming level results in a better framework for comparing the timing of these abrupt shifts. We determined the global warming level in each model by calculating the difference between the 10-year smoothed yearly mean time series of near-surface air temperature from the 1pctCO$_2$ and the pre-industrial control simulations.

## 3. Results

### 3.1. Global Analysis of Abrupt Shifts in CMIP6

Figure 2 shows a global overview of detected abrupt shifts in the CMIP6 model ensemble. The heat map in Figure 2a denotes the fraction of models that have at least one abrupt shift per grid cell in any month in any variable. Certain climate subsystems experience more abrupt shifts than others; especially the polar regions exhibit many abrupt shifts on the level of grid cells. The boxes indicate the different climate subsystems we have analyzed in this study (except for the monsoons, see figure caption). See Table 1 for an overview of these systems, their coordinates, key variables, and area thresholds we have chosen to define large-scale abrupt shifts in each subsystem.

To give an impression of how these large-scale abrupt shifts look like, example time series of these shifts are shown in Figure 4 for each climate subsystem. By aggregating the large-scale abrupt shifts from all models and all key variables in each subsystem, we obtained a distribution of the global warming level at which these shifts occur. Figure 2b shows these distributions per climate subsystem together with the estimated temperature thresholds by Armstrong McKay et al. (2022). The orange numbers show the proportion of models with at least one large-scale abrupt shift per subsystem.

There is a large difference in the number of abrupt shifts between CMIP6 models. There are nine models without any large-scale abrupt shifts in any of the climate subsystems (large-scale: abrupt shifts covering at least the area given in Table 1 for each subsystem). 48 models have at least one large-scale abrupt shift in any subsystem. The majority of models (39) have between 2 and 4 subsystems with a large-scale abrupt shift. There are three models with five or more subsystems exhibiting large-scale abrupt shifts. Figure S5 and Table S3 in Supporting Information S1 summarize the subsystems and variables with abrupt shifts per model.

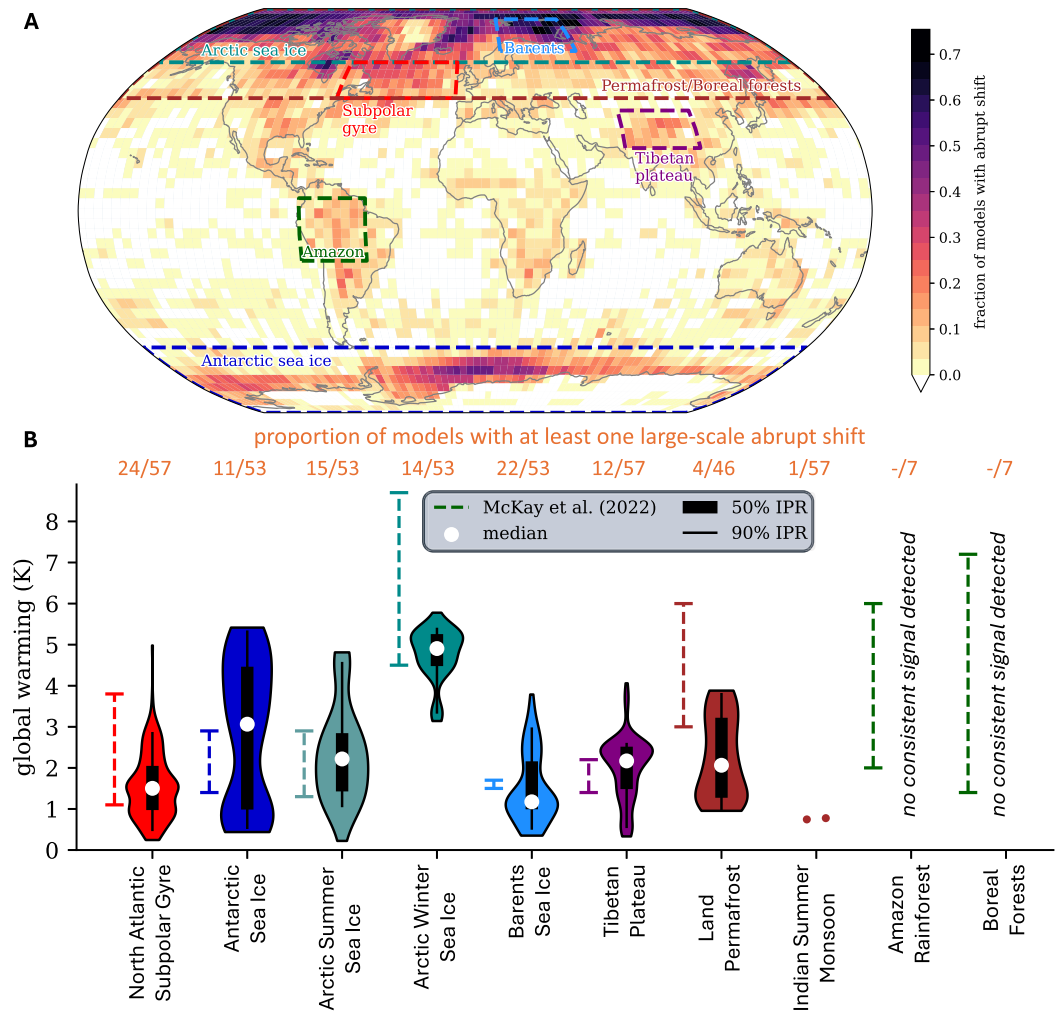

**Figure 2.** (a) Fraction of models that have an abrupt shift per grid cell in at least one of the 82 tested variables. This is based on the abrupt shifts at a grid cell level (the results from individual simulations are overlaid on a common 4 by 4° grid). Highlighted are the different systems of focus in this paper; tipping elements: North Atlantic subpolar gyre, Tibetan Plateau, Land permafrost, Amazon rainforest, Boreal forests; unclear tipping elements: Antarctic sea ice, Monsoons; not tipping elements: Arctic summer sea ice, Arctic winter sea ice, Barents sea ice. The boxes for the monsoons are not drawn, but they include the South American Monsoon, West African Monsoon, and Indian Summer Monsoon. (b) The mean global temperature increase at which large-scale abrupt shifts are detected in different climate subsystems. Each violin plot illustrates the normalized density distribution of global warming at which large-scale abrupt shifts are detected across all models for all key variables relevant to each subsystem (see Table 1 for the key variables). A large-scale abrupt shift is defined as one covering a surface area exceeding a specific threshold, as detailed in Table 1 for each subsystem. The white dots show the mean, the thick black lines show the 50% intercentile range (IPR), and the thin black lines the 90% IPR. For the Indian Summer Monsoon only two instances of abrupt shifts are found, these are denoted by the two dots. The orange numbers indicate the proportion of models with large-scale abrupt shift among those models with key variables per subsystem. The dashed lines show the global warming thresholds as estimated by McKay et al. (Armstrong McKay et al., 2022) if available.

To test whether the number of abrupt shifts in a model depends on its climate sensitivity, we performed two-sided Pearson correlation tests between the number of subsystems with large-scale abrupt shifts per model with both the models' Transient Climate Response (TCR) and Equilibrium Climate Sensitivity (ECS) (values taken from Intergovernmental Panel On Climate Change (2023); Hausfather et al. (2022)). TCR is defined as the level of global warming after 70 years from a 1% annual increase in $CO_2$ starting from preindustrial conditions until $CO_2$ doubling. ECS represents the global warming increase resulting from a doubling of $CO_2$ until an equilibrium is reached. We excluded the Amazon rainforest and Boreal forests from this analysis, as only 7 out of 57 models include dynamic vegetation components, and no consistent signals of abrupt shifts were detected (see the Amazon Rainforest and Boreal Forests sections). The correlation between TCR and the number of subsystems with large-scale abrupt shifts was $r = 0.532$ ($p < 0.001$), while the correlation with ECS was $r = 0.625$ ($p < 0.001$), both indicating a significant, moderate, and positive correlation. Thus, models with higher warming rate generally have more large-scale abrupt shifts than those with lower warming rates.

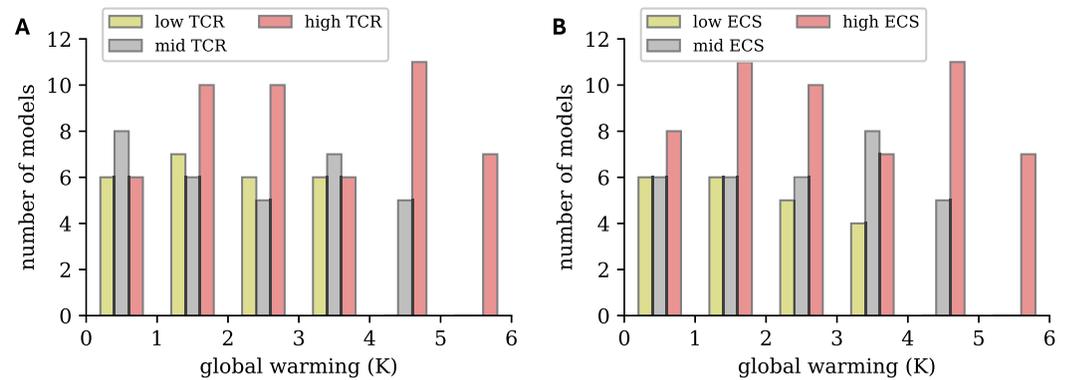

**Figure 3.** Number of models that show a large-scale abrupt shift across global warming bins. Panel A shows the number of models with a large-scale abrupt shift in any of the subsystems (excluding the Amazon rainforest and the Boreal forests), grouped by global warming levels in 1 K intervals. The models are divided into three equal-size groups based on their Transient Climate Response (TCR): low TCR (1.33–1.77 K), mid TCR (1.80–2.05 K), and high TCR (2.06–2.99 K). Panel B shows the same distribution, but groups the models by their Equilibrium Climate Sensitivity (ECS): low ECS (1.83–3.04 K), mid ECS (3.05–4.28 K), and high ECS (4.31–5.63 K).

These positive correlations can be explained by two factors. First, models with higher climate sensitivity (higher TCR/ECS) have faster warming rates, which can lead to more abrupt shifts due to more rapidly changing variables. Second, these models reach higher global temperatures by the end of the 1pctCO₂ scenario than models with low climate sensitivity, making abrupt shifts more likely in subsystems with high temperature thresholds, such as Antarctic sea ice or Arctic winter sea ice.

Figure 3 supports this, showing the number of models with large-scale abrupt shifts binned by global warming levels in 1° K intervals, with models divided into three equal-sized groups of low, mid, and high TCR (Figure 3a) and ECS (Figure 3b). At lower warming levels, models with high climate sensitivity generally have more abrupt shifts. Additionally, many of these models show abrupt shifts at global warming levels between 4 and 6 K. These temperatures are not reached by models with lower climate sensitivity.

Many of the CMIP6 models are structurally related. They often share components or are derived from common models, which can lead to correlated outputs. To assess whether these structural families differ in the number of subsystems with large-scale abrupt shifts (again excluding the Amazon rainforest and Boreal forests), we conducted a nonparametric Kruskal-Wallis test. The models were divided into structural families based on Kuma et al. (2023). Given the highly unequal family sizes (ranging from 1 to 20 models), we excluded families with fewer than three models to increase the reliability of the test. The test showed no statistically significant difference in the number of subsystems with abrupt shifts between the structural families (H-statistic of 5.55, $p = 0.475$).

In the next subsections, we analyze each climate subsystem individually. We first look at subsystems identified as tipping elements, then at subsystems identified as unclear tipping elements, and last at subsystems identified as not tipping elements. The code of the analysis is available via Terpstra et al. (2025). This repository also contains figures showing the time series and spatial extent of all detected large-scale abrupt shifts in key variables of all subsystems.

### 3.2. Subsystems Identified as Tipping Elements

#### 3.2.1. North Atlantic Subpolar Gyre

In the North Atlantic subpolar gyre, ocean convection occurs. The sinking of dense waters in this region contributes to the AMOC. The subpolar gyre is identified as a separate tipping element from the AMOC and a collapse of convection is expected to occur abruptly with medium confidence (Lenton, 2023). Observations also indicate reduced warming over the subpolar gyre (Drijfhout et al., 2012), potentially related to a weakening of convection in this region. As evidence for a collapse, we expect to find negative abrupt shifts—defined as abrupt declines—in the sea surface salinity, the mixed-layer depth, the sea surface temperature and—related to the latter —the surface air temperature (Sgubin et al., 2017; Swingedouw et al., 2021).

In total, 24 out of 57 models show an abrupt shift in the subpolar gyre for the sea surface salinity, the mixed-layer depth, the sea surface temperature, or the surface air temperature. The surface air temperature has been used in a previous study to identify abrupt shifts in the subpolar gyre (Swingedouw et al., 2021). For this variable we find seven models with an abrupt shift, which occur between 0.31 and 2.03 K of global warming (90% IPR (interpercentile range)). For sea surface temperature, 22 models show an abrupt shift. All but one (CNRM-CM6-1-HR) of the models with an abrupt shift in surface air temperature also show an abrupt cooling in sea surface temperature. There are thus more abrupt shifts in sea surface temperature than in surface air temperature. This difference may result from a damping effect, where the cooling of the atmosphere is less pronounced compared to the cooling of the ocean surface. The sea surface temperature shifts occur between 0.49 and 2.99 K of global warming (90% IPR).

For observing convection, the mixed-layer depth is the most suitable variable because it provides a more direct measure of convection than surface temperatures. Eleven models show an abrupt shift in the mixed-layer depth, of which 10 also exhibit an abrupt cooling of sea surface temperature (only IPSL-CM5A2_INCA does not). The abrupt shifts in mixed-layer depth occur between 0.47 and 2.14 K of global warming (90% IPR). On average they cover at most a few grid cells, which is to be expected since convection is a local process. Figure 4a shows an example of such an abrupt shift in the mixed-layer depth in MRI-ESM2-0. Lastly, for sea surface salinity, 2 models show an abrupt shift, both over reasonably large areas. These models are BCC-ESM1 (occurring at 0.97 K of global warming, with the area being at least $5.7 \times 10^5$ km$^2$) and CESM2-FV2 (occurring at 1.35 K of global warming, with the area being at least $9.8 \times 10^5$ km$^2$). The locations of the abrupt shifts do not happen at consistent locations and seem highly model-dependent.

In CMIP5, a few models were found to show abrupt shifts in the Labrador sea related to convection (Drijfhout et al., 2015; Sgubin et al., 2017). Here, we identify 19 models as having abrupt shifts in the subpolar gyre. The risk of abrupt shifts in the subpolar gyre region has also been assessed by Swingedouw et al. (2021) for CMIP6, where they considered the average surface air temperature over a box covering the subpolar gyre region instead of grid cell data as is done here. For 2 of the 3 models for which they identify abrupt shifts (CESM2-WACCM, MRI-ESM2-0), we also detect abrupt shifts. Using the edge detection method no abrupt shifts are found for the third model (NorESM2-LM). This can be due to the different model scenarios used (here 1pctCO$_2$, Swingedouw et al. (2021) used the Shared Socio-Economic Pathways (O'Neill et al., 2016; Riahi et al., 2017)). The high global mean temperature increase in the 1pctCO$_2$ scenario offsets part of the cooling caused by a subpolar gyre collapse, which could lead to missing some events in sea surface and surface air temperatures. Still, we detect more abrupt shifts in the sea surface temperature than in the mixed-layer depth, likely because the high variability in the latter makes it difficult to distinguish abrupt shifts from natural fluctuations.

#### 3.2.2. Tibetan Plateau

Glaciers outside the large polar ice masses can exhibit complicated nonlinear dynamics, mostly driven by (local) surface air temperature changes. Therefore, the world's glaciers including those on the Tibetan Plateau have been classified as regional tipping elements (medium confidence) (Lenton, 2023). While glacier dynamics are not resolved in the CMIP6 models considered here, the snow cover on such glaciers can be assessed. In CMIP5, two models experienced abrupt decline of the snow cover in the Tibetan Plateau region (Drijfhout et al., 2015).

In 12 out of 57 CMIP6 models we find a large-scale abrupt shift in at least one key variable (see Table 1). The surface upwelling shortwave radiation, which is affected by the surface albedo, shows the highest proportion of abrupt shifts, with shifts occurring in 11 models between 0.69 and 2.60 K of global warming (90% IPR). An abrupt decline in albedo is most likely caused by the disappearance of snow cover.

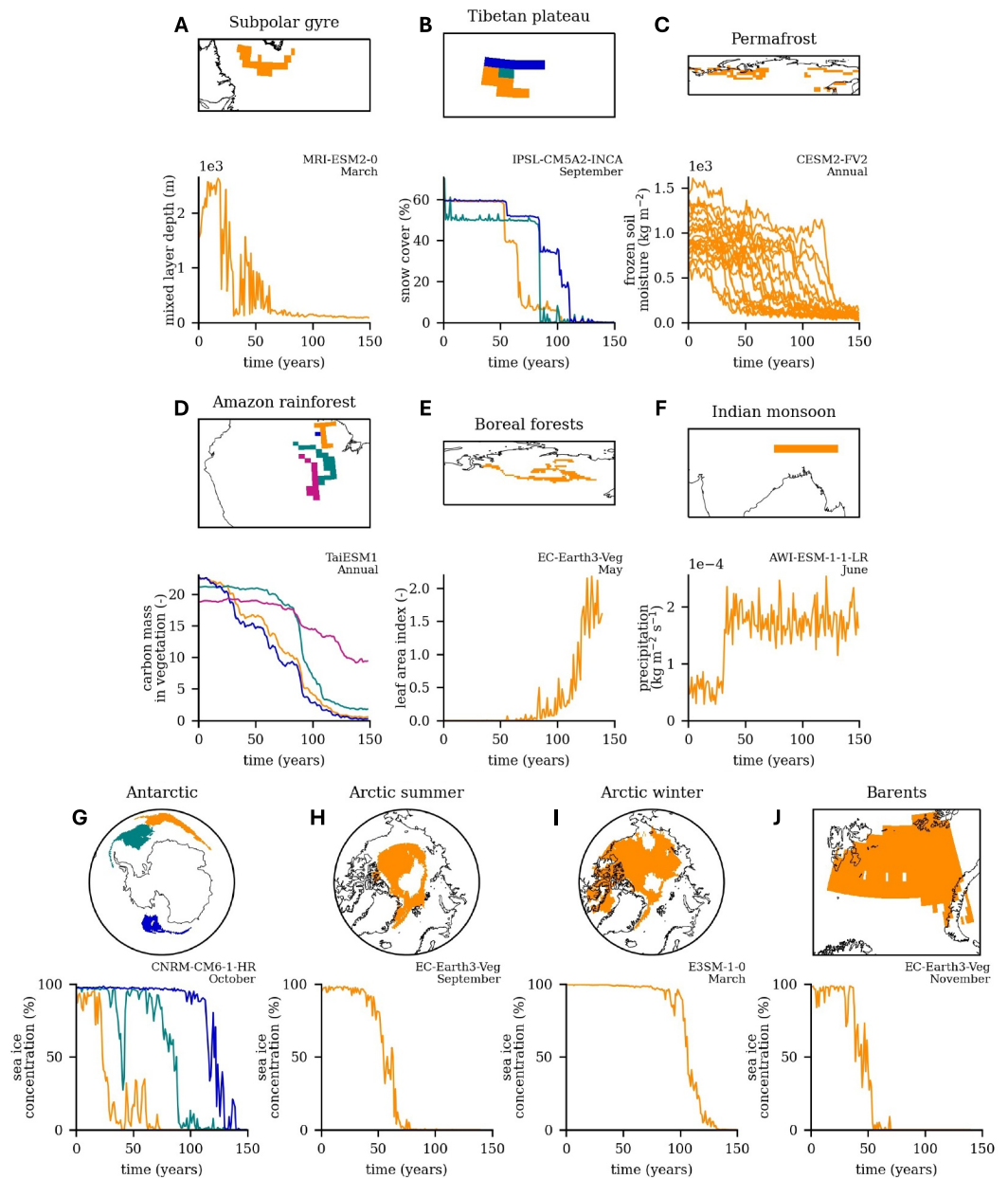

**Figure 4.** The spatial extent and time series of a large-scale abrupt shift for a representative variable and model is shown for each of (a) North Atlantic subpolar gyre, (b) Tibetan Plateau, (c) Amazon rainforest, (d) Boreal forests, (e) Land permafrost, (f) Indian summer Monsoon, (g) Antarctic sea ice, (h) Arctic summer sea ice, (i) Arctic winter sea ice, and (j) Barents sea ice. In each panel, the model and month in which the abrupt shift(s) occur are denoted (where "Annual" indicates a yearly average instead). The spatial plots are located within the boxes of Figure 2a. The Boreal forest and land permafrost subsystems are zoomed into regions of Northern Asia. The Indian summer monsoon example is located south of the Tibetan plateau.

Three variables directly represent the snow cover of the Tibetan Plateau: snow area percentage, surface snow melt, and snow depth. Of these, abrupt shifts in snow depth occur at the lowest level of global warming between 0.36 and 2.61 K (90% IPR) in 7 models. Abrupt shifts in snow area percentage occur between 1.41 and 3.59 K (90% IPR, 5 models) and for surface snow melt between 1.90 and 3.82 K (90% IPR, 2 models). The models with abrupt shifts in these last two variables are subsets of the models with detected shifts in snow depth. This indicates that snow depth declines abruptly before the surface snow melts (since the abrupt shifts in snow depth occur at a lower level of global warming than in snow area percentage and surface snow melt). Finally, we find 4 models (of which 3 are versions of the CanESM5 model) that have an abrupt shift in near-surface air temperature between 1.19 and 2.52 K of global warming (90% IPR). Figure 4b shows the abrupt decline of snow area percentage in IPSL-CM5A2-INCA.

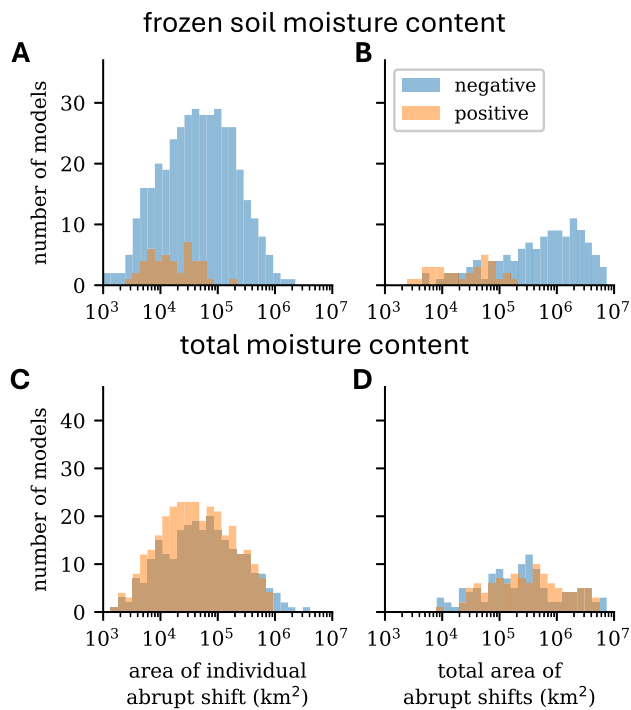

**Figure 5.** Number of models showing abrupt shifts in frozen soil water content and total soil moisture content for land permafrost, depending on the area they cover. Panels (a) and (c) show the number of models as a function of the area covered by individual abrupt shifts in any month, distinguishing between positive (increase, orange) and negative (decrease, blue) shifts for frozen soil moisture content and total soil moisture content, respectively. Panels (b) and (d) shows a similar distribution, but now for the total area affected by these shifts in each month for frozen soil moisture content and total soil moisture content, respectively.

Overall, only 12 models out of 57 show an abrupt shift in snow-related variables in the CMIP6 models spread over a limited area. However, since glacier dynamics are not resolved in the CMIP6 models, the number of abrupt shifts in the Tibetan Plateau might be underestimated because abrupt shifts in the glaciers themselves cannot be represented.

### 3.2.3. Land Permafrost

In the GTP-report (Lenton, 2023), the land permafrost areas in the Northern Hemisphere are classified as regional tipping elements (medium confidence), while, on a global scale, permafrost-related processes are not considered to have tipping potential with medium confidence. Abrupt thaw of permafrost occurs through processes such as thermokarst formation, leading to subsiding ground and rapid expansion of thaw lakes (Grosse et al., 2013). Without these nonlinear processes, permafrost thaw is considered more gradual. On a global scale, the positive carbon-climate feedback (extra $CO_2$ and methane emissions from soil degradation after thaw) could make land permafrost a tipping element. However, this feedback is currently considered too small to be important globally (Lenton, 2023), and the simulations considered here do not include it because they are driven by prescribed $CO_2$ concentrations. Despite this, Drijfhout et al. (2015) did find one CMIP5 model exhibiting an abrupt change in total soil moisture content, indicating an abrupt melt of permafrost in the Arctic tundra. Apart from the total soil moisture content, we focus our analysis on the soil frozen water content, as this is a proxy for permafrost (Steinert et al., 2023).

Both total soil moisture content and soil frozen water content show abrupt shifts in the CMIP6 models. However, there is a large difference between the two variables in the fraction of positive abrupt shifts (i.e., abrupt increase) and negative abrupt shifts (i.e., abrupt decrease). Figure 5 shows the number of positive and negative abrupt shifts detected over all models for frozen soil moisture content (panel A) and total soil moisture content (panel C) depending on the area these shifts cover. For total soil moisture content, there are many positive as well as negative shifts (Figure 5c). However, for abrupt shifts covering more than 1 million km²;, there are only negative ones with 4 out of 46 models exhibiting a large-scale negative abrupt shift (between 0.95 and 3.84 K of global warming (90% IPR)).

In contrast to total soil moisture content, soil frozen water content shows a considerably higher number of negative (i.e., abrupt decline) than positive abrupt shifts (i.e., abrupt increase), most of which are small shifts (Figure 5a). Just one model shows large-scale negative abrupt shifts covering an area of at least 1 million km² each (UKESM1-0-LL). There are 36 separate instances of abrupt shifts in this model that occur between 1.17 and 3.27 K of global warming (90% IPR). Figure 4c shows all abrupt declines in CESM2-FV2 in Northern Asia for the annual time series for soil frozen water content.

However, since permafrost is expected to be a regional tipping point, we have also looked at the total area of all abrupt shifts in a simulation (i.e., the total area of abrupt shifts in a single month or year) which is displayed in Figures 5b and 5d for both positive and negative shifts. For frozen soil moisture, there are 19 out of 36 models with negative abrupt shifts covering a total area of at least 1 million km², with the largest total area being $6.3 \times 10^6$ km². For total soil moisture content, in 8 models the abrupt shifts cover a total area of at least 1 million km².

In summary, we find abrupt shifts in frozen soil moisture content. However, individual abrupt shifts are small (covering an area of less than 1 million km$^2$) except for 1 model. Taking into account the total area of abrupt shifts in a single simulation, we find that, with 19 models, permafrost can still disappear abruptly on a regional scale.

### 3.2.4. Amazon Rainforest

The Amazon rainforest is considered a tipping element by the GTP-report (Lenton, [2023]). The report evaluates the tipping potential of this element with high confidence at the local scale, medium confidence at the regional scale, and low confidence at the continental scale. Shifts at all three scales are expected to occur abruptly with medium confidence.

In our assessment, we only consider the seven CMIP6 models that have dynamic vegetation. These are EC-Earth3-Veg, GFDL-ESM4, MPI-ESM1-2-LR, NorCMP1, TaiESM1, SAM0-UNICON, and UKESM1-0-LL. Among these, EC-Earth-Veg, GFDL-ESM4, MPI-ESM1-2-LR, and NorCMP1 also simulate fire. We detect many abrupt shifts, most of which are localized events. However, some of the considered variables, such as carbon mass, have low interannual variability, such that the computed abruptness measure detects abrupt shifts that are very small in magnitude. Moreover, the timescale of vegetation dynamics is generally slower (decades) than in the subsystems we analyzed previously, which is why the edge detector is not optimally configured.

Abrupt shifts within CMIP6 models were also analyzed by Parry et al. ([2022]) under the 1pctCO$_2$ scenario. They scanned for abrupt shifts in carbon mass in vegetation using specific criteria: the vegetation carbon must change by at least 2 kg C m$^{-2}$ over a 15-year period, contributing to at least 25% of the overall change, with the mean annual rate of change being at least three times larger than the variability in the control simulation. Their analysis revealed abrupt shifts in all models except for UKESM1-0-LL. These abrupt shifts were mostly localized but added up to between 0% and 40% of total area in the northern South America region of the Amazon at 3 K of global warming.

When we apply a similar additional criterion, requiring at least a 25% increase or decrease up to 30 years before and after the abrupt shift (using the same parts of the time series as for the abruptness measure, see Section [2]) we still detect abrupt shifts in bare soil percentage area coverage (1 model), carbon mass in vegetation (3 models), total carbon mass flux from vegetation to litter (1 model), carbon mass flux out of atmosphere due to gross primary production on land (3 models), natural grass area percentage (2 models), leaf area index (3 models), total atmospheric respiration on land as carbon mass flux (2 models), and tree cover percentage (2 models). These abrupt shifts occur between 0.85 and 4.95 K of global warming (90% IPR), and regard relatively small areas. Figure [4d] shows an example of abrupt shifts in TaiESM1.

### 3.2.5. Boreal Forests

In the GTP-report (Lenton, [2023]), (Northern Hemisphere) Boreal forest changes have been classified as tipping elements on the regional scale (dieback on the southern edge with medium confidence, expansion on the Northern edge with low confidence). As for the Amazon rainforest, we assess abrupt shifts in the subset of CMIP6 simulations that include dynamic vegetation. Suggested drivers of such abrupt changes are regional drying and atmospheric warming, which could potentially be related to permafrost thaw. Southern dieback refers to a tipping point that leads to an almost treeless state (steppe/prairie), while Northern expansion could be an abrupt expansion of tree cover into previously tundra or little vegetated areas. In Drijfhout et al. ([2015]), the abrupt northward expansion of Boreal forests has been observed in two CMIP5 models.

Although considered abrupt, forest expansion still occurs on timescales of more than a decade, making it challenging to detect with the edge detector within the short time series of the 1pctCO$_2$ scenario and the chosen parameters of the edge detector as discussed in the previous section on the Amazon rainforest. We apply the same extra condition (at least 25% increase or decrease up to 30 years period before and after the abrupt shift) here, and find abrupt shifts in bare soil percentage area coverage (2 models), carbon mass in vegetation (2 models), total carbon mass flux from vegetation to litter (1 model), total carbon mass flux from vegetation to soil (1 model), carbon mass flux out of atmosphere due to gross primary production on land (5 models), natural grass area percentage (3 models), leaf area index (4 models), total atmospheric respiration on land as carbon mass flux (2 models), and tree cover percentage (2 models). These abrupt shifts occur between 0.80 and 4.87 K of global warming (90% IPR) and include mostly northward expansion. Figure 4e shows an example of an abrupt increase of leaf area index for EC-Earth3-Veg.

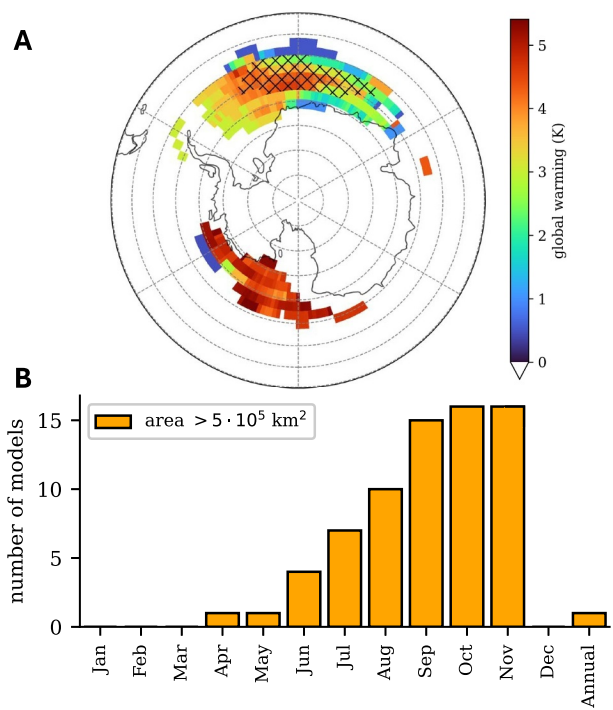

**Figure 6.** Abrupt shifts in Antarctic sea ice. (a) The spatial distribution of the global warming level at which large-scale abrupt shifts occur in Antarctic sea ice concentration. The hashed overlay indicates that in that grid cell, at least 5 models show an abrupt shift (the results from individual simulations are overlaid on a common 4 by 4° grid) There is one cluster located around the Weddell Sea, and one around the Amundsen and Ross Seas. The first cluster generally experiences abrupt shifts at lower levels of global warming than the second cluster. (b) The number of models that have at least one abrupt shift larger than the given area threshold for each month.

Overall, we find some abrupt shifts occurring in the Boreal forests, however, more targeted analysis is necessary to quantify and understand these transitions in CMIP6 models in more detail.

### 3.3. Subsystems Identified as Unclear Tipping Elements

#### 3.3.1. Monsoons

In the GTP-report (Lenton, 2023), only the West African Monsoon is categorized as a tipping element, but with low confidence. For the Indian Summer Monsoon and the South American Monsoon it is unclear whether they are tipping elements. For the Indian Summer Monsoon transitions have been found in proxy data, that is cave records (Gupta et al., 2019), where large changes in precipitation are associated with changing monsoon intensity. The suggested main mechanism behind possible tipping is the moisture-advection feedback, where increased precipitation leads to increased latent heating, causing an increased land-ocean temperature difference and hence landward advection of moisture, amplifying the precipitation perturbation (Levermann et al., 2009). We focus the analysis on latent heat flux and both general and convective precipitation, which are linked to this mechanism.

We find no abrupt shifts in (convective) precipitation related to the South American Monsoon and West African Monsoon. For the Indian Summer Monsoon, we find two models with abrupt shifts in precipitation. In one model, AWI-ESM-1-1-LR, an abrupt increase in precipitation is detected in both June and August (see Figure 4f). The abrupt shifts are located close to the Himalaya (covering an area of 1.5 and 1.7 x $10^5$ km$^2$ for the 2 months). This might be a slight northward shift of the Indian Monsoon system. In the other model, FGOALS-g3, we detect an abrupt increase in precipitation off the coast of south-east India near the last 20 years of the simulation (see Figure S6 in Supporting Information S1). Upon visual inspection, it is not clear whether this is an extreme event or a persisting abrupt shift since the precipitation seems to return to its baseline. There are no abrupt shifts detected in latent heat flux.

Overall, we find very limited evidence of abrupt shifts in the Indian Summer Monsoon system and no evidence of abrupt shifts in the West African Monsoon and the South American Monsoon systems.

#### 3.3.2. Antarctic Sea Ice

Antarctic sea ice reaches its minimum extent in February and its maximum extent in September. The GTP-report (Lenton, 2023) classifies Antarctic sea ice as an unclear tipping element, also stating that it is uncertain whether it undergoes abrupt shifts.

We found that 36 out of 53 CMIP6 models show an abrupt shift in sea ice concentration of which only 3 models (CNRM-CM6-1-HR, GISS-E2-1-H, UKESM1-1-LL) show an abrupt shift covering more than 1 million km$^2$ (the time series for CNRM-CM6-1-HR is shown in Figure 4g). These 3 models show a large spread in the timing of the shifts, ranging between 0.51 and 4.67 K of global warming (90% IPR). Large-scale abrupt shifts covering an area of at least 0.5 million km$^2$ are more numerous, with 11 models displaying such a shift. Figure 6b shows that all of them—except for one in each of April, May, and Annal—occur between June and November. No large-scale abrupt shifts are observed from December to March, likely because of relatively little sea ice in summer in the Antarctic. GISS-E2-1-H is the only model to show an abrupt shift in April and May (occurring at 1 K of global warming). The level of global warming at which the abrupt shifts occur from June to November ranges between 0.54 and 5.41 K (90% IPR).

The detected large-scale abrupt shifts in sea ice concentration occur in two different locations, as Figure 6a illustrates. One is around the Weddell Sea and the other around the Amundsen and Ross Seas. In the Weddell Sea the abrupt shifts in sea ice concentration occur between 0.58 and 5.23 K of global warming (90% IPR), and around the Amundsen and Ross Sea between 2.51 and 5.41 K. The abrupt shifts in the Weddell Sea not only occur at lower levels of global warming but also shows more abrupt shifts in sea ice concentration than the Amundsen and Ross Sea, occurring in 10 models compared to 5.

### 3.4. Subsystems Not Identified as Tipping Elements

#### 3.4.1. Arctic Summer and Winter Sea Ice

Arctic sea ice is often divided according to the season into summer and winter sea ice. Arctic sea ice coverage reaches its minimum extent in September. The GTP-report (Lenton, 2023) classifies Arctic summer sea ice as neither a tipping element nor a system that can undergo an abrupt shift, with high confidence. This is because the melting of Arctic summer sea ice is thought to be a threshold-free process, lacking a self-reinforcing mechanism that accelerates melting beyond control once it crosses a critical threshold. Arctic sea ice coverage reaches its maximum extent in March. Winter sea ice is classified as not being a tipping element with medium confidence by the GTP-report (Lenton, 2023). However, it is expected with high confidence, in contrast to the Arctic summer sea ice, that a part of the Arctic winter sea ice can disappear abruptly due to the geometry of ice thickness (Bathiany et al., 2016). The abrupt decline of Arctic winter sea ice can also be explained by positive feedback loops between atmosphere and ocean (Hankel & Tziperman, 2021).

For the summer sea ice, we analyzed August, September and October, since the minimum extent is seen in September (Shu et al., 2020), while for the winter, we analyzed February, March and April, since the maximum extent is in March (Shu et al., 2020). We took 2 months around the standard months to report summer and winter sea ice since the maximum and minimum might vary slightly depending on the particular model and the global warming level. In the Arctic summer sea ice, 44 out of 53 CMIP6 models show an abrupt shift, with 15 models displaying a large-scale abrupt shift. For sea ice concentration, 11 models show large-scale abrupt shifts with the onset between 1.12 and 3.13 K of global warming (90% IPR) (see Figure 4h for an example). In the Arctic winter sea ice, 48 out of 53 CMIP6 models show an abrupt shift, with 14 models displaying a large-scale abrupt shift. In sea ice concentration, 13 models show large-scale abrupt shifts, with the onset between 3.41 and 5.50 K of global warming (90% IPR) (see Figure 4i for an example).

When taking into account the other scanned variables (see Table 1), as also displayed in Figure 2b, the onset of large-scale abrupt shifts occurs between 1.04 and 4.58 K of global warming (90% IPR) for summer and between 3.28 and 5.41 K of global warming (90% IPR) for winter. For summer, 3 models have a large-scale abrupt shift in near-surface air temperature occurring between 1.49 and 4.76 K of global warming (90% IPR), which is higher than the interval for sea ice concentration. However, these 3 models do not show large-scale abrupt shifts in sea ice concentration.

Figure 7a shows the spatial distribution of the onset of large-scale abrupt shifts in terms of global warming for Arctic summer sea ice. Shown per grid cell is the average level of global warming at which these abrupt shifts occur in different models. Figure 7b shows the same but for Arctic winter sea ice. The abrupt shifts in winter occur over a larger area than in summer, which is to be expected since winter sea ice has a larger extent than summer sea ice.

When considering all months of the year, not only do we see a clear difference between summer and winter, but also one between the occurrence of abrupt shifts at the minimum versus maximum extent of sea ice. Figure 7c shows that from August until November there is a lower onset of abrupt shifts than in December until June. This division corresponds to those months belonging to the group of minimum extent versus maximum extent of sea ice (below or above the average maximum and minimum extent in historical context in CMIP6 models (Shu et al., 2020)). For the minimum extent group, large-scale abrupt shifts occur between 1.16 and 3.40 K of global warming (90% IPR), while for the maximum extent group large-scale abrupt shifts occur between 3.09 and 5.40 K of global warming (90% IPR).

In their analysis, Drijfhout et al. (2015) found 4 models where sea ice disappeared abruptly over an area larger than 4 million km². We find that 7 models (CIESM, CMCC-CM2-SR5, CMCC-ESM2, E3SM-1-0, FIO-ESM-2-0, NESM3, and UKESM1-1-LL) have such an abrupt shift (which only occurs in January until May) with onsets between 3.21 and 5.24 K of global warming (90% IPR). Altogether, in partial contrast to the assessment of the GTP-report (Lenton, 2023) (where they state that abrupt shifts are not expected in Arctic summer sea ice), the CMIP6 models show that not only the Arctic winter sea ice, but also the Arctic summer sea ice might undergo abrupt shifts over a substantial part of the Arctic region. Whether this is indeed due to Arctic geometry, needs further study.

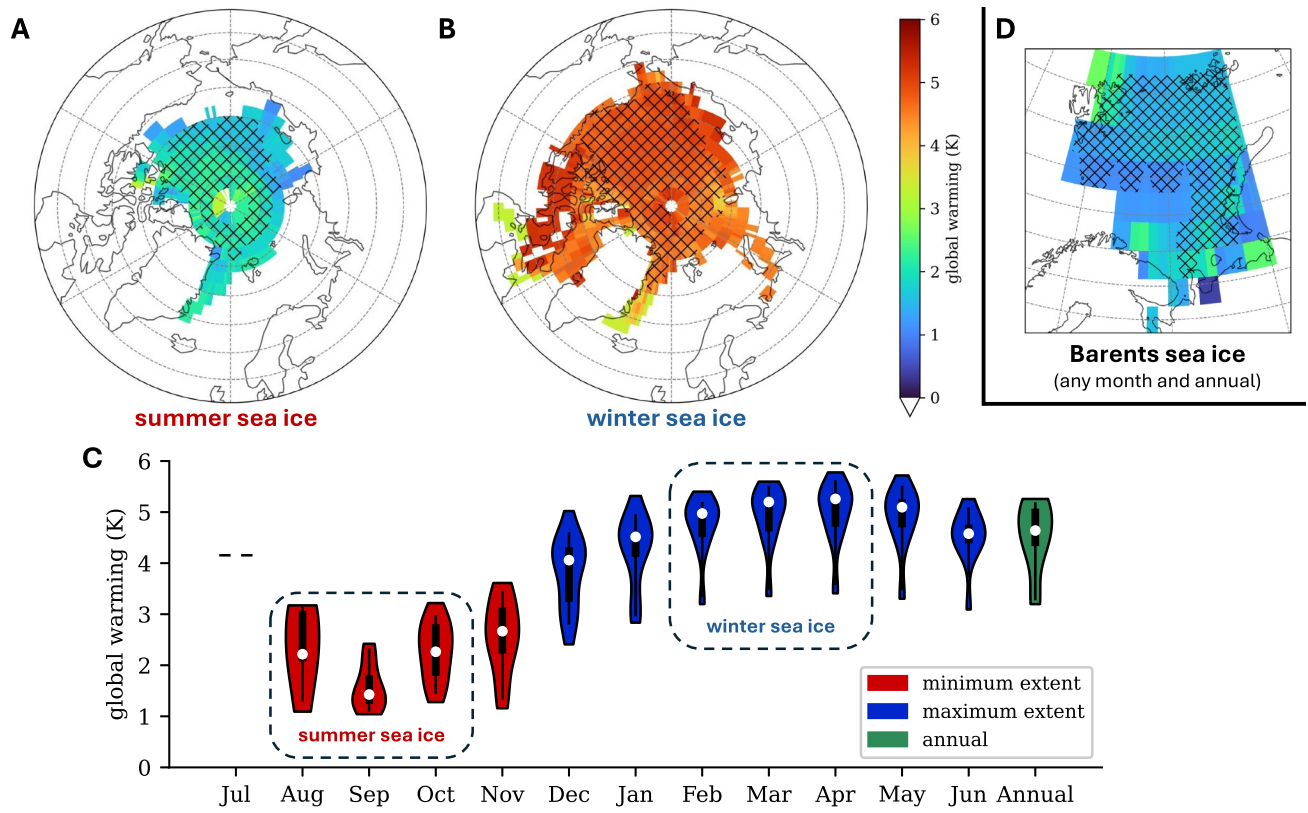

**Figure 7.** Spatial and temporal occurrence of abrupt shifts of Arctic sea ice. (a) The spatial distribution of the global warming level at which large-scale abrupt shifts occur in Arctic summer sea ice (defined as sea ice extent in August, September, and October), (b) the spatial distribution for Arctic winter sea ice (defined as sea ice extent in February, March, and April), and (d) the spatial distribution for Barents sea ice. The hashed overlays indicates that in that grid cell at least 5 models show an abrupt shift (the results from individual simulations are overlaid on a common 4 by 4° grid). (c) The distribution of global warming levels at which large-scale abrupt shifts occur in the Arctic sea ice per month. The red violin plots correspond to sea ice of minimum extent, the blue violin plots show the sea ice of maximum extent, and the green violin plot the annual sea ice. There is only one model with a large-scale abrupt shift in July. The white dots show the mean, the thick black lines show the 50% interpercentile range, and the thin black lines show the 90% interpercentile range.

### 3.4.2. Barents Sea Ice

Barents sea ice is mostly winter sea ice located above Scandinavia and Western Russia and is therefore a subset of the overall Arctic sea ice. It is not classified as a tipping element, albeit with low confidence due to limited studies on tipping of this system (Lenton, 2023).

We found that 43 out of 53 models show an abrupt shift in the Barents Sea. Of these, 18 show a large-scale abrupt shift in sea ice concentration (covering an area over 0.2 million km$^2$; Figure S7 in Supporting Information S1). These large-scale shifts occur between 0.49 and 2.85 K of global warming (90% IPR). There are two models with abrupt shifts covering more than 0.5 million km$^2$: EC-Earth3-Veg and EC-Earth3-Veg-LR occurring between 1.16 and 1.30 K of global warming (see Figure 4j for EC-Earth3-Veg time series). Figure 7d shows the spatial distribution of global warming levels at which the large-scale abrupt shifts occur in sea ice concentration. The abrupt shifts are most prevalent in the northern and eastern part and less toward the south. This is likely because not all models have sea ice extended close to Northern Scandinavia.

In summary, we find evidence that Barents sea ice can undergo abrupt shifts according to the CMIP6 models. Some of the shifts we find here are part of larger shifts in Arctic sea ice since they belong to larger connected components. We only considered the parts of the bigger components that fell within the Barents Sea.

## 4. Discussion

We assessed the presence of abrupt shifts in the CMIP6 model ensemble under a scenario of strongly increasing $CO_2$ concentrations. We used edge detection, a method to identify and quantify abrupt shifts in various climate subsystems. The majority of models (48 out of 57) have a large-scale abrupt shift in at least one of the studied systems. There is a large diversity in how many abrupt shifts the models show, indicating that some models are more prone to abrupt shifts than others. We detected at least one abrupt shift in all climate subsystems we examined, although the number of occurrences of these shifts vary greatly between subsystems.

In the sea ice systems, we detected more shifts in the Arctic than in the Antarctic region. 11 models had a large-scale abrupt shift in Antarctic sea ice concentration occurring between 0.55 and 5.41 K (90% IPR) of global warming. This is a wider range than reported by Armstrong McKay et al. (2022), namely a range of 1.4–2.9 K.

The GTP-report (Lenton, 2023) states that Arctic summer sea ice is not a tipping element, nor is it expected to disappear abruptly. However, we found that large-scale abrupt shifts do occur in Arctic summer sea ice. For the sea ice concentration, 17 models have a large-scale abrupt shift occurring between 1.12 and 3.13 K (90% IPR) of global warming. This range is similar to the estimated range of 1.3–2.9 K by Armstrong McKay et al. (2022). However, when including all key variables, our range is between 1.04 and 4.58 K (90% IPR). This increase in range is largely caused by the abrupt shifts in near-surface air temperature and surface downwelling radiation which occur at relatively high levels of global warming in the cases we detected. The abrupt shifts in Arctic sea ice are possibly due to the increase in natural variability once sea ice retreats, thus observing synchronized retreats of sea ice due to internal variability (Eisenman, 2010) without it being a tipping point.

For the Arctic winter sea ice concentration, 21 models have a large-scale abrupt shift between 3.41 and 5.50 K (90% IPR) of global warming. This range is lower than 4.5–8.7 K as estimated by Armstrong McKay et al. (2022). One possible reason for the lower upper bound of the temperature thresholds is that most CMIP6 models do not reach a global warming level of 8.7 K in the 1pctCO$_2$ scenario. Including all key variables, this range shifts only a little to 3.28 and 5.41 K (90% IPR). Finally, 18 models have a large-scale abrupt shift for the Barents sea ice concentration, occurring between 0.49 and 2.85 K (90% IPR) of global warming which is a wider range than estimated by Armstrong McKay et al. (2022) who place it between 1.5 and 1.7 K.

In the two analyzed vegetation systems, the Amazon rainforest and the Boreal forest, we detected abrupt shifts in various variables. However, the edge detection, with the settings we used, performed less well on vegetation-related variables. This is due to both the low variability in many of these variables and that vegetation tends to respond over longer timescales than the decade-long timescale used with the edge detection. In general, we found numerous, disconnected, abrupt shifts occurring over a wide range of global warming levels ranging between 0.80 and 4.87 K (90% IPR).

In the Tibetan Plateau, we detected abrupt shifts in snow-related variables in several CMIP6 models. These shifts are most prevalent in surface upwelling shortwave radiation, a proxy for albedo, indicating a sudden loss of snow cover, with 12 models showing an abrupt shift. Additionally, we found abrupt shifts in snow area percentage (4 models), melt rate (2 models), and snow depth (6 models). Drijfhout et al. (2015) also identified two CMIP5 models with an abrupt decline in snow cover of the Tibetan Plateau. The level of global warming at which all these shifts occur in CMIP6 is between 0.53 and 2.61 K (90% IPR). This range overlaps with the global warming threshold estimated by Armstrong McKay et al. (2022), who place it between 1.4 and 2.2 K of global warming, although we also detect abrupt shifts below their lower bound of 1.4 K. Abrupt shifts in glaciers could not be assessed because they are typically not dynamically represented in CMIP models. Nevertheless, the presence of abrupt shifts in snow cover is clear in both CMIP5 and CMIP6 models.

For the land permafrost, we found more abrupt shifts than Drijfhout et al. (2015) who found only one: an abrupt shift in total soil moisture content in HadGEM2-ES. However, the CMIP6 variant of this model was excluded from our assessment due to not having the simulation output available at the time of analysis (see Text S1.C in Supporting Information S1). The abrupt shifts detected in CMIP6, although generally individually small, add up to larger areas. For soil frozen water content, only one model has an individual abrupt shift covering more than 1 million km$^2$. However, 19 models have a total area of abrupt shifts of more than 1 million km$^2$. These occur between 0.63 and 4.59 K of global warming. This is a lower range than the global warming threshold estimated by Armstrong McKay et al. (2022) who place it between 3.0 and 6.0 K of global warming.

In the monsoon systems, we only found abrupt shifts in one model for the Indian Summer Monsoon and none for the South American and West African Monsoons. For the Indian summer monsoon, previous work (Katzenberger et al., 2021) suggests linearly increasing rainfall with rising global temperatures under the SSP5-8.5 scenario in CMIP6 models (when averaged across the entire monsoon region). Similarly, no abrupt shifts in the monsoons were found in the CMIP5 ensemble (Drijfhout et al., 2015). Since we detected only one model with (local) abrupt shifts in just one of the monsoon systems, we consider abrupt shifts unlikely to occur in the monsoons based on the CMIP6 models.

Overall, the results support previous assessments indicating that multiple climate subsystems might undergo abrupt shifts (Armstrong McKay et al., 2022; Bathiany et al., 2020; Drijfhout et al., 2015; Lenton, 2023). However, the level of global warming at which these shifts occur remains highly uncertain. Notably, we find the lower bound of the global warming thresholds for the North Atlantic subpolar gyre, Tibetan Plateau, Antarctic sea ice, and Arctic winter sea ice is lower in our assessment than in that of Armstrong McKay et al. (2022). It is important to note that their assessment was based not only on CMIP models, but consisting of multiple lines of evidence from the literature, and they assessed for tipping and not just abrupt shifts as in this study. Table S4 in Supporting Information S1 provides an overview of the global warming thresholds from our assessment, Drijfhout et al. (2015), and Armstrong McKay et al. (2022).

The high uncertainty in global warming levels at which abrupt shifts occur in CMIP6 models has multiple causes. One contributing factor is the grouping of shifts across different months. In some systems, abrupt shifts can occur in different months and at different levels of global warming, often depending on the season during which the shift takes place. This is evident with the Arctic sea ice, where there is a significant difference in the onset of abrupt shifts between summer and winter. However, seasonal variation alone does not fully account for the substantial differences between models. Other contributing factors include the different ways each model resolves the underlying physics and the diversity in climate sensitivity across the models.

In addition to the variation in the timing of abrupt shifts across different models and variables, this study's estimates of global warming levels at which these shifts occur might be overestimated. The 1pctCO$_2$ scenario represents a situation of rapidly increasing CO$_2$ concentrations, whereas the current CO$_2$ concentration increases more slowly. Due to this rapid increase of CO$_2$ in the simulations, the edge detector may identify abrupt shifts at higher warming levels than it would under a more gradual increase, as systems require time to adjust to a rising CO$_2$ concentration. However, the high rate of warming might also trigger rate-dependent tipping, where the system is forced into a new state due to the rate of forcing. As a result, some estimates of the global warming required to trigger abrupt shifts might be underestimated compared to those from scenarios with slower forcing rates, more closely aligned with the current rate of global warming.

To detect the abrupt shifts, we made several methodological choices, though alternative approaches are possible. Most importantly, the definition of the threshold for the spatial extent of large-scale abrupt shifts remains somewhat subjective, even though we based it on the specific characteristics of each subsystem. The field lacks an objective definition of how large an abrupt shift must be to be considered relevant since its impact can be measured from multiple perspectives—for example, its effect on other climate subsystems, dependent ecosystems, or socio-economic systems.

Detecting abrupt shifts in a systematic way remains challenging given the extensive and diverse model outputs from CMIP6. Here, we use edge detection, a general tool that can be applied to all systems without requiring system-specific knowledge. However, designing a measure to quantify abruptness while avoiding detecting abrupt shifts of small magnitude remains challenging, as different variables exhibit different noise characteristics and shifts occur on different scales.

Overall, we find evidence of abrupt shifts in many climate subsystems within CMIP6. The number and onset of abrupt shifts vary significantly between models. Still, these results should be interpreted as not including all potential abrupt shifts, given that the 1pctCO$_2$ simulations used are relatively short (150 years) and exclude some important tipping elements, such as ice sheets. Longer simulations that include all climate subsystems are needed for a more accurate assessment of the long-term evolution and likelihood of abrupt shifts in tipping elements under climate change.

Despite aforementioned limitations, our analysis shows that higher levels of global warming increase the risk of abrupt shifts across the studied models. Moreover—with the exception of the monsoon systems—we found abrupt shifts in all systems across multiple models. The climate subsystems that we included in our analysis and were previously shown to have abrupt shifts in CMIP5, also showed abrupt shifts in CMIP6 (Drijfhout et al., 2015). Many of the detected abrupt shifts occur only over parts of a subsystem and does not always imply a full collapse of a complete subsystem. Our estimates of uncertainty ranges of the timing of abrupt shifts generally overlap with the estimates of Armstrong McKay et al. (2022), although it is important to note that they look at tipping points and not just abrupt shifts in each system as done in this study. Importantly, even at a global warming of 1.5 K, six out of the 10 subsystems studied here show a risk of large-scale abrupt shifts.

## Acronyms

| | |
|---|---|
| 1pctCO$_2$ | CMIP6 simulation with a 1 percent per year increase in CO$_2$, starting from pre-industrial CO$_2$ concentrations |
| AMOC | Atlantic Meridional Overturning Circulation |
| CMIP6 | Coupled Model Intercomparison Project Phase 6 |
| ECS | Equilibrium Climate Sensitivity |
| GTP-report | Global Tipping Points Report (Lenton, 2023) |
| IPR | Interpercentile range |
| TCR | Transient Climate Response |

## Conflict of Interest

The authors declare no conflicts of interest relevant to this study.

## Data Availability Statement

The code of the analysis is available via Terpstra et al. (2025). We used publicly available data which can be obtained from the CMIP6 archive of the Earth System Grid Federation (ESGF) (Cinquini et al., 2014). With the provided scripts and this data, all results of this manuscript can be reproduced. The code repository also contains the edge data of all detected large-scale abrupt shifts and all other data necessary to easily reproduce the figures in this manuscript. Finally, the repository contains figures showing the time series and spatial extent of all detected large-scale abrupt shifts in key variables of all subsystems.

**Acknowledgments**

This publication is part of the project "Interacting climate tipping elements: When does tipping cause tipping?" (with project number VI.C.202.081 of the NWO Talent programme) financed by the Dutch Research Council (NWO). This is ClimTip contribution #15; the ClimTip project has received funding from the European Union's Horizon Europe research and innovation programme under grant agreement No. 101137601.

## References From the Supporting Information